\documentclass[journal]{IEEEtran}

\usepackage{color}
\usepackage{float}
\usepackage{graphicx}
\usepackage{epstopdf}  
\usepackage{psfrag}
\usepackage{epsfig}
\usepackage[FIGTOPCAP]{subfigure}
\usepackage{mathtools}
\usepackage[utf8]{inputenc}
\usepackage{gensymb}
\usepackage{cite}

\begin{document}

\title{Efficient Analysis of Metasurfaces in Terms of Spectral-Domain GSTC Integral Equations}

\author{Nima~Chamanara,
        Karim~Achouri,~\IEEEmembership{Student Member,~IEEE,}
        Christophe~Caloz,~\IEEEmembership{Fellow,~IEEE}
\thanks{Polytechnique Montr\'{e}al, Montr\'{e}al, Qu\'{e}bec H3T 1J4, Canada.}
}


\maketitle

\begin{abstract}
We present a spectral-domain (SD) technique for the efficient analysis of metasurfaces. The metasurface is modeled by generalized sheet transition conditions (GSTCs) as a zero-thickness sheet creating a discontinuity in the electromagnetic field. The SD expression of these GSTCs for a specified incident field leads to a system of four surface integral equations for the reflected and transmitted fields, which are solved using the method of moments in the spectral domain. Compared to the finite-difference and finite-element techniques that require meshing the entire computational domain, the proposed technique reduces the problem to the surface of the metasurface, hence eliminating one dimension and providing substantial benefits in terms of memory and speed. A monochromatic generalized-refractive metasurface and a polychromatic focusing metasurface are presented as illustrative examples.
\end{abstract}


%
\IEEEpeerreviewmaketitle

\section{Introduction}


Metasurfaces are 2D arrangements of scattering particles that transform electromagnetic waves in specified ways. In contrast to 3D metamaterials, they are easier to fabricate, and compared to frequency selective surfaces (FSSs), they provide richer transformations. Metasurfaces have already found a great diversity of applications, including polarization transformers and rotators~\cite{niemi2013synthesis,kodera2011artificial}, perfect absorbers~\cite{ra2013total}, nonreciprocal screens~\cite{kodera2011artificial,Sounas_TAP_01_2013,hadad2015space,taravati2016nonreciprocal}, beam transformers~\cite{salem2014manipulating}, electromagnetic interferometers and processors~\cite{achouri2016metasurface}.

A metasurface can be effectively modeled as a zero-thickness sheet that creates a discontinuity in the electromagnetic field. Such a discontinuity can be most generally described by generalized sheet transition conditions (GSTCs)~\cite{idemen1987boundary, achouri2014general, achouri2015synthesis, chamanara2016spacetime}, where the discontinuity in the electric and magnetic fields are related to surface electric and magnetic current and surface electric and magnetic polarization densities. GSTCs are not implemented in currently available commercial software and their future implementation will require efficient numerical approaches. Exact zero-thickness modeling by GSTCs has been recently incorporated in academic numerical solvers based on finite differences~\cite{vahabzadeh2016simulation}, including finite-difference time-domain (FDTD)~\cite{taflove2005computational}, finite-difference frequency-domain (FDFD)~\cite{chamanara2013_opex_isol}, and finite-element methods~\cite{sandeep2016finite, jin2015finite}. The FDTD-GSTC and FDFD-GSTC techniques~\cite{vahabzadeh2016simulation} properly model the zero thickness of the metasurface through the introduction of virtual nodes at the location of the metasurface on, while the FEM-GSTC does this naturally through flexible FEM mesh points at both sides of the metasurface.

Although they model the zero thickness electromagnetic field discontinuity quite well, the aforementioned GSTC finite-difference and finite-element techniques require to mesh the entire computational domain. In large problems, and particularly three-dimensional ones, volumetric meshes represent a high computational burden in terms of memory and speed. In contrast, surface integral equations reduce the problem complexity by one dimension, and are hence able to solve large problems that may be intractable with other techniques. As GSTCs describe the electromagnetic discontinuity on a surface, it is natural to cast them to the form of surface integral equations~\cite{harrington1996field, ishimaru1991electromagnetic} and thus essentially reduce the problem to the surface of the metasurface. Moreover, for a flat metasurface, GSTCs may be conveniently expressed in the spectral domain (SD)~\cite{itoh1989numerical} and analyzed with the well developed Fourier techniques~\cite{itoh1989numerical}. This paper combines these two points to provide an exact analysis of metasurfaces in the spectral domain (SD)~\cite{itoh1989numerical} based on SD integral equations. GSTCs are modeled as exact zero-thickness transition conditions and transformed into a system of SD surface integral equations which are then solved using the method of moments (MoM)~\cite{harrington1996field}. The integral equations are derived and solved in the temporal frequency domain. However, the method can be generalized to time domain SD integral equation as well if the problem is linear.

The organization of the paper is as follows. GSTCs are described in Sec.~\ref{sec:Gstc}. SD integral equations are derived in Sec.~\ref{sec:SDA}. MoM solution of the SD integral equations are presented in Sec.~\ref{sec:MoM}. Two illustrative examples are provided in Sec.~\ref{sec:example}. Finally, Sec.~\ref{sec:conclusion} concludes the paper.

\section{Transition Conditions} \label{sec:Gstc}

\begin{figure}[ht!]
\centering
\psfrag{a}[l][c][1.0]{$\mathbf{E}^\text{i}, \mathbf{H}^\text{i}$}
\psfrag{b}[l][c][1.0]{$\mathbf{E}^\text{t}, \mathbf{H}^\text{t}$}
\psfrag{c}[l][c][1.0]{$\mathbf{E}^\text{r}, \mathbf{H}^\text{r}$}
\psfrag{x}[l][c][0.9]{$x$}
\psfrag{y}[l][c][0.9]{$y$}
\psfrag{z}[l][c][0.9]{$z$}
\includegraphics[page=1, width=0.9\columnwidth]{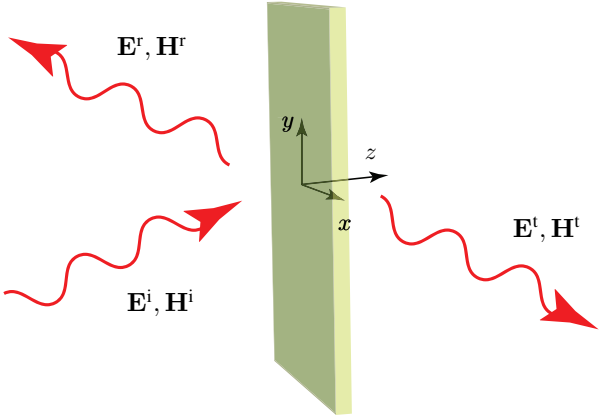}
\caption{General metasurface transformation: a given incident electromagnetic wave is transformed into specified reflected and transmitted waves. The metasurface is assumed to be position in the $xy$-plane at $z=0$ of cartesian coordinate system.}
\label{fig:gstc-transform}
\end{figure}

Figure~\ref{fig:gstc-transform} represents a general metasurface transformation, where a given incident wave is transformed into specified reflected and transmitted waves. Assuming that the metasurface is placed in the $xy$-plane at $z=0$, the electromagnetic field discontinuity it induces may be rigorously modeled by the GSTCs~\cite{idemen1987boundary,achouri2015synthesis}
\begin{subequations}
\begin{align}
\hat{\mathbf{z}}\times\Delta\mathbf{H}\left(\boldsymbol{\rho},t\right)&=\frac{\partial}{\partial t}\mathbf{P}_{T}\left(\boldsymbol{\rho},t\right)-\hat{\mathbf{z}}\times\nabla_T M_{z}\left(\boldsymbol{\rho},t\right),\\
\mathbf{\hat{z}}\times\Delta\mathbf{E}\left(\boldsymbol{\rho},t\right)&=-\mu_{0}\frac{\partial}{\partial t}\mathbf{M}_{T}\left(\boldsymbol{\rho},t\right)-\frac{1}{\epsilon_{0}}\hat{\mathbf{z}}\times\nabla_T P_{z}\left(\boldsymbol{\rho},t\right),
\end{align} \label{eq:GSTCs}
\end{subequations}
\noindent
where $\mathbf{P}$ and $\mathbf{M}$ represent the generally space-time dependent electric and magnetic surface polarization densities, and where
\begin{subequations}\label{eq:diff_fields}
\begin{align}
\Delta\mathbf{E}\left(\boldsymbol{\rho},t\right)&=\mathbf{E}^{+}\left(\boldsymbol{\rho},t\right)-\mathbf{E}^{-}\left(\boldsymbol{\rho},t\right),\\ \Delta\mathbf{H}\left(\boldsymbol{\rho},t\right)&=\mathbf{H}^{+}\left(\boldsymbol{\rho},t\right)-\mathbf{H}^{-}\left(\boldsymbol{\rho},t\right),
 \end{align}
\end{subequations}
with superscripts $^{+}$ and $^{-}$ representing the total fields on the right and left of the metasurface, respectively, subscripts $_T$ and $_z$ representing the components tangential and normal to metasurface, respectively, and $\boldsymbol{\rho}$ representing an arbitrary point on the metasurface, i.e. $\boldsymbol{\rho}=x\hat{\mathbf{x}}+y\hat{\mathbf{y}}$.

Assuming that the metasurface is dispersive but linear and local, the polarization densities are related to the average ($_\text{av}$)  electric and magnetic fields on the metasurface through the general bianisotropic dispersion relations
\begin{subequations}
\begin{align}
\mathbf{P}\left(\boldsymbol{\rho},\omega\right) =& \epsilon_0 \bar{\bar{\boldsymbol{\chi}}}_\text{ee}\left(\boldsymbol{\rho},\omega\right)\cdot\mathbf{E}_\text{av}\left(\boldsymbol{\rho},\omega\right) \\
&+ \sqrt{\mu_0\epsilon_0}\bar{\bar{\boldsymbol{\chi}}}_\text{em}\left(\boldsymbol{\rho},\omega\right)\cdot \mathbf{H}_\text{av}\left(\boldsymbol{\rho},\omega\right), \nonumber\\
\mathbf{M}\left(\boldsymbol{\rho},\omega\right) =& \sqrt{\frac{\epsilon_0}{\mu_0}}\bar{\bar{\boldsymbol{\chi}}}_\text{me}\left(\boldsymbol{\rho},\omega\right)\cdot \mathbf{E}_\text{av}\left(\boldsymbol{\rho},\omega\right) \\
&+ \bar{\bar{\boldsymbol{\chi}}}_\text{mm}\left(\boldsymbol{\rho},\omega\right)\cdot \mathbf{H}_\text{av}\left(\boldsymbol{\rho},\omega\right), \nonumber
\end{align}
\end{subequations}
\noindent
with
\begin{subequations}\label{eq:av_fields}
\begin{align}
\mathbf{E}_\text{av}\left(\boldsymbol{\rho},\omega\right)&=\left[\mathbf{E}^+\left(\boldsymbol{\rho},\omega\right)+\mathbf{E}^-\left(\boldsymbol{\rho},\omega\right)\right]/2,\\ \mathbf{H}_\text{av}\left(\boldsymbol{\rho},\omega\right)&=\left[\mathbf{H}^+\left(\boldsymbol{\rho},\omega\right)+\mathbf{H}^-\left(\boldsymbol{\rho},\omega\right)\right]/2, \end{align}
\end{subequations}
where $\omega$ represents the temporal frequency. The functions of $\left(\boldsymbol{\rho},\omega\right)$ in these relations are related to their temporal Fourier transforms $\left(\boldsymbol{\rho},t\right)$ functions by the Fourier transform pair

\begin{subequations}
\begin{align}
\boldsymbol{\Psi}\left(\boldsymbol{\rho},t\right)=& \int_{-\infty}^{+\infty}\boldsymbol{\Psi}\left(\boldsymbol{\rho},\omega\right)e^{j\omega t}d\omega, \\
\boldsymbol{\Psi}\left(\boldsymbol{\rho},\omega\right)=& \frac{1}{2\pi}\int_{-\infty}^{+\infty}\boldsymbol{\Psi}\left(\boldsymbol{\rho},t\right)e^{-j\omega t}dt,
\end{align}
\end{subequations}

\noindent where $\boldsymbol{\Psi}=\mathbf{P}$, $\mathbf{M}$, $\mathbf{E}$ and $\mathbf{H}$. The next section will derive the GSTCs in the spatiotemporal $(\mathbf{k},\omega)$ spectral domain and develop the corresponding SD integral equations for the reflected and transmitted fields.

\section{Spectral-Domain Analysis}	\label{sec:SDA}

It is assumed that incident field and metasurface susceptibilities are known. This section derives the SD integral equations for the corresponding reflected and transmitted fields. The known incident field and the unknown reflected and transmitted fields are decomposed into their spatiotemporal Fourier components $(\mathbf{k}_T,\omega)$ as follows:
\begin{subequations}
\begin{align}
\mathbf{E}^\text{i}\left(\mathbf{r},t\right)&=\int\tilde{\mathbf{E}}^\text{i}\left(\mathbf{k}_{T},\omega\right)e^{j(\omega t-\mathbf{k}_T\cdot\boldsymbol{\rho} - k_zz)}d\omega d\mathbf{k}_{T}, \\
\mathbf{E}^\text{r}\left(\mathbf{r},t\right)&=\int\tilde{\mathbf{E}}^\text{r}\left(\mathbf{k}_{T},\omega\right)e^{j(\omega t-\mathbf{k}_T\cdot\boldsymbol{\rho} + k_zz)}d\omega d\mathbf{k}_{T}, \\
\mathbf{E}^\text{t}\left(\mathbf{r},t\right)&=\int\tilde{\mathbf{E}}^\text{t}\left(\mathbf{k}_{T},\omega\right)e^{j(\omega t-\mathbf{k}_T\cdot\boldsymbol{\rho} - k_zz)}d\omega d\mathbf{k}_{T},
\end{align} \label{eq:E-kw}
\end{subequations}
\noindent where $\mathbf{k}_T=k_x\hat{\mathbf{x}} + k_y\hat{\mathbf{y}}$ and where the integral sign specifically represent the triple integral
\begin{equation}
\int\cdots d\omega d\mathbf{k}_{T} = \int_{-\infty}^{+\infty}\int_{-\infty}^{+\infty}\int_{-\infty}^{+\infty}\cdots d\omega dk_xdk_y.
\end{equation}
\noindent
With these definitions, the differential operators $\nabla$ and $\nabla_T$ reduce to their algebraic SD equivalents
\begin{subequations}
\begin{align}
\nabla &\rightarrow -j(\mathbf{k}_T \pm k_z\hat{\mathbf{z}}), \\
\nabla_T &\rightarrow -j\mathbf{k}_T,
\end{align}
\end{subequations}
where $k_z$ is found as
\begin{equation}
k_z = \sqrt{k_0^2 - k_x^2 - k_y^2},
\end{equation}
with $k_0=\omega_0\sqrt{\epsilon_0\mu_0}$, upon inserting~\eqref{eq:E-kw} into the free-space Helmholtz equation, \mbox{$(\nabla^2+k_0^2)\mathbf{E}\left(\mathbf{r},\omega\right)=0$}, and decomposing the result $k^2+k_0^2=0$ into transverse and normal components. The sign of the square root must be chosen so as to satisfy the radiation condition at $z=\pm\infty$, i.e. $\Re(k_z)\ge 0$, $\Im(k_z)\le 0$.

The magnetic fields corresponding to~\eqref{eq:E-kw} are obtained through the Maxwell-Faraday equation, $\mathbf{H}\left(\mathbf{r},\omega\right) = \frac{-j\omega}{\mu_0}\nabla\times\mathbf{E}\left(\mathbf{r},\omega\right)$, as
\begin{subequations}
\begin{align}
\mathbf{H}^\text{i}\left(\mathbf{r},t\right)=\frac{1}{\omega\mu_0}\int&\left(\mathbf{k}_{T}+k_{z}\hat{\mathbf{z}}\right)\times\\
&\tilde{\mathbf{E}}^\text{i}\left(\mathbf{k}_{T},\omega\right)e^{j(\omega t-\mathbf{k}_T\cdot\boldsymbol{\rho} - k_zz)}d\omega d\mathbf{k}_{T} \nonumber,\\
\mathbf{H}^\text{r}\left(\mathbf{r},t\right)=\frac{1}{\omega\mu_0}\int&\left(\mathbf{k}_{T}-k_{z}\hat{\mathbf{z}}\right)\times\\
&\tilde{\mathbf{E}}^\text{r}\left(\mathbf{k}_{T},\omega\right)e^{(j\omega t-\mathbf{k}_T\cdot\boldsymbol{\rho} + k_zz)}d\omega d\mathbf{k}_{T} \nonumber,\\
\mathbf{H}^\text{t}\left(\mathbf{r},t\right)=\frac{1}{\omega\mu_0}\int&\left(\mathbf{k}_{T}+k_{z}\hat{\mathbf{z}}\right)\times\\
&\tilde{\mathbf{E}}^\text{t}\left(\mathbf{k}_{T},\omega\right)e^{j(\omega t-\mathbf{k}_T\cdot\boldsymbol{\rho} - k_zz)}d\omega d\mathbf{k}_{T}. \nonumber
\end{align} \label{H-kw}
\end{subequations}

We next assume, for simplicity and without loss of generality, monoanisotropic (rather than bianisotropic) metasurfaces, i.e.
\begin{subequations}
\begin{align}
\mathbf{P}\left(\boldsymbol{\rho},\omega\right)&=\varepsilon_{0}\mathbf{\boldsymbol{\bar{\bar{\chi}}}}_\text{ee}\left(\boldsymbol{\rho},\omega\right)\mathbf{\mathbf{E}_\text{av}}\left(\boldsymbol{\rho},\omega\right), \\
\mathbf{M}\left(\boldsymbol{\rho},\omega\right)&=\mathbf{\boldsymbol{\bar{\bar{\chi}}}}_\text{mm}\left(\boldsymbol{\rho},\omega\right)\mathbf{H}_\text{av}\left(\boldsymbol{\rho},\omega\right),
\end{align} \label{eq:mono-aniso-r}
\end{subequations}
\noindent
where the electric and magnetic susceptibility tensors are also expanded in the spectral domain:

\begin{subequations}
\begin{align}
\mathbf{\bar{\bar{\boldsymbol{\chi}}}}_\text{ee}\left(\boldsymbol{\rho},t\right)&=\int\tilde{\bar{\bar{\boldsymbol{\chi}}}}_\text{ee}\left(\mathbf{k}_{T},\omega\right)e^{j(\omega t-\mathbf{k}_T\cdot\boldsymbol{\rho})}d\omega d\mathbf{k}_{T}, \\
\mathbf{\bar{\bar{\boldsymbol{\chi}}}}_\text{mm}\left(\boldsymbol{\rho},t\right)&=\int\tilde{\bar{\bar{\boldsymbol{\chi}}}}_\text{mm}\left(\mathbf{k}_{T},\omega\right)e^{j(\omega t-\mathbf{k}_T\cdot\boldsymbol{\rho})}d\omega d\mathbf{k}_{T}.
\end{align}
\end{subequations}

\noindent In the spatial-frequency Fourier domain the product expressions in~\eqref{eq:mono-aniso-r} become convolutions, and the corresponding transverse and normal polarization densities read thus
\begin{subequations}
\begin{align}
\mathbf{\tilde{P}}_{T}\left(\mathbf{k}_{T},\omega\right)=\varepsilon_{0}\int\tilde{\bar{\bar{\boldsymbol{\chi}}}}_\text{ee}^{T}\left(\mathbf{k}_{T}^{\prime},\omega\right)\mathbf{{\tilde{\mathbf{E}}_\text{av}}}\left(\mathbf{k}_{T}-\mathbf{k}_{T}^{\prime},\omega\right)d\mathbf{k}_{T}^{\prime}, \\
\mathbf{\tilde{M}}_{T}\left(\mathbf{k}_{T},\omega\right)=\int\tilde{\bar{\bar{\boldsymbol{\chi}}}}_\text{mm}^{T}\left(\mathbf{k}_{T}^{\prime},\omega\right)\mathbf{\mathbf{\tilde{H}}_\text{av}}\left(\mathbf{k}_{T}-\mathbf{k}_{T}^{\prime},\omega\right)d\mathbf{k}_{T}^{\prime}.
\end{align} \label{eq:pm-T-EH-kw}
\end{subequations}
and
\begin{subequations}
\begin{align}
\tilde{P}_{z}\left(\mathbf{k}_{T},\omega\right)=\varepsilon_{0}\int\tilde{\bar{\bar{\boldsymbol{\chi}}}}_\text{ee}^{z}\left(\mathbf{k}_{T}^{\prime},\omega\right)\mathbf{\mathbf{\tilde{E}}_\text{av}}\left(\mathbf{k}_{T}-\mathbf{k}_{T}^{\prime},\omega\right)d\mathbf{k}_{T}^{\prime}, \\
\tilde{M}_{z}\left(\mathbf{k}_{T},\omega\right)=\int\tilde{\bar{\bar{\boldsymbol{\chi}}}}_\text{mm}^{z}\left(\mathbf{k}_{T}^{\prime},\omega\right)\mathbf{\mathbf{\tilde{H}}_\text{av}}\left(\mathbf{k}_{T}-\mathbf{k}_{T}^{\prime},\omega\right)d\mathbf{k}_{T}^{\prime}.
\end{align} \label{eq:pm-z-EH-kw}
\end{subequations}
where

\begin{subequations}
\begin{align}
\tilde{\bar{\bar{\boldsymbol{\chi}}}}_\text{ee}^{T}\left(\mathbf{k}_{T},\omega\right) &= \\
&\left[\begin{array}{ccc}
\chi_{ee}^{xx}\left(\mathbf{k}_{T},\omega\right) & \chi_{ee}^{xy}\left(\mathbf{k}_{T},\omega\right) & \chi_{ee}^{xz}\left(\mathbf{k}_{T},\omega\right)\\
\chi_{ee}^{yx}\left(\mathbf{k}_{T},\omega\right) & \chi_{ee}^{yy}\left(\mathbf{k}_{T},\omega\right) & \chi_{ee}^{yz}\left(\mathbf{k}_{T},\omega\right)
\end{array}\right], \nonumber\\
\tilde{\bar{\bar{\boldsymbol{\chi}}}}_\text{mm}^{T}\left(\mathbf{k}_{T},\omega\right) &= \\
&\left[\begin{array}{ccc}
\chi_\text{mm}^{xx}\left(\mathbf{k}_{T},\omega\right) & \chi_\text{mm}^{xy}\left(\mathbf{k}_{T},\omega\right) & \chi_\text{mm}^{xz}\left(\mathbf{k}_{T},\omega\right)\\
\chi_\text{mm}^{yx}\left(\mathbf{k}_{T},\omega\right) & \chi_\text{mm}^{yy}\left(\mathbf{k}_{T},\omega\right) & \chi_\text{mm}^{yz}\left(\mathbf{k}_{T},\omega\right)
\end{array}\right], \nonumber\\
\tilde{\bar{\bar{\boldsymbol{\chi}}}}_\text{ee}^{z}\left(\mathbf{k}_{T},\omega\right) &= \\
&\left[\begin{array}{ccc}
\chi_{ee}^{zx}\left(\mathbf{k}_{T},\omega\right) & \chi_{ee}^{zy}\left(\mathbf{k}_{T},\omega\right) & \chi_{ee}^{zz}\left(\mathbf{k}_{T},\omega\right)
\end{array}\right], \nonumber\\
\tilde{\bar{\bar{\boldsymbol{\chi}}}}_\text{mm}^{z}\left(\mathbf{k}_{T},\omega\right) &= \\
&\left[\begin{array}{ccc}
\chi_\text{mm}^{zx}\left(\mathbf{k}_{T},\omega\right) & \chi_\text{mm}^{zy}\left(\mathbf{k}_{T},\omega\right) & \chi_\text{mm}^{zz}\left(\mathbf{k}_{T},\omega\right)
\end{array}\right]. \nonumber
\end{align}
\end{subequations}

Finally, the terms involving $\hat{\mathbf{z}}\times\nabla_T$ in~\eqref{eq:GSTCs} become in SD form

\begin{subequations}
\begin{align}
\mathbf{z}\times\nabla_T P_{z}\left(\boldsymbol{\rho},t\right)=\int\hat{\mathbf{z}}\times\mathbf{k}_{T}\tilde{P}_{z}\left(\mathbf{k}_{T},\omega\right)e^{j(\omega t-\mathbf{k}_T\cdot\boldsymbol{\rho})}d\omega d\mathbf{k}_{T}, \\
\mathbf{z}\times\nabla_T M_{z}\left(\boldsymbol{\rho},t\right)=\int\hat{\mathbf{z}}\times\mathbf{k}_{T}\tilde{M}_{z}\left(\mathbf{k}_{T},\omega\right)e^{j(\omega t-\mathbf{k}_T\cdot\boldsymbol{\rho})}d\omega d\mathbf{k}_{T}.
\end{align} \label{eq:z-nablaT-pm-z-kw}
\end{subequations}

At this point, substituting~\eqref{eq:E-kw}, \eqref{H-kw}, \eqref{eq:pm-T-EH-kw} and \eqref{eq:z-nablaT-pm-z-kw} into~\eqref{eq:GSTCs}, and using the orthogonality property of Fourier harmonics, yields the following SD integral equations:
\begin{figure*}[!ht]
\begin{equation}
\begin{split}
&\frac{1}{\omega\mu_0}\hat{\mathbf{z}}\times\left(-\left(\mathbf{k}_{T}+k_{z}\hat{\mathbf{z}}\right)\times\tilde{\mathbf{E}}^\text{i}\left(\mathbf{k}_{T},\omega\right)-\left(\mathbf{k}_{T}-k_{z}\hat{\mathbf{z}}\right)\times\tilde{\mathbf{E}}^\text{r}\left(\mathbf{k}_{T},\omega\right)+\left(\mathbf{k}_{T}+k_{z}\hat{\mathbf{z}}\right)\times\tilde{\mathbf{E}}^\text{t}\left(\mathbf{k}_{T},\omega\right)\right)= \\
&\frac{1}{2}i\omega\varepsilon_{0}\int\tilde{\bar{\bar{\boldsymbol{\chi}}}}_\text{ee}^{T}\left(\mathbf{k}_{T}-\mathbf{k}_{T}^{\prime},\omega\right)\left[\tilde{\mathbf{E}}^\text{i}\left(\mathbf{k}_{T}^{\prime},\omega\right)+\tilde{\mathbf{E}}^\text{r}\left(\mathbf{k}_{T}^{\prime},\omega\right)+\tilde{\mathbf{E}}^\text{t}\left(\mathbf{k}_{T}^{\prime},\omega\right)\right]d\mathbf{k}_{T}^{\prime}\\
&-\frac{1}{2\omega\mu_0}\int\hat{\mathbf{z}}\times\left(\mathbf{k}_{T}^{\prime}-\mathbf{k}_{T}\right)\tilde{\bar{\bar{\boldsymbol{\chi}}}}_\text{mm}^{z}\left(\mathbf{k}_{T}^{\prime}-\mathbf{k}_{T},\omega\right)
\left[\left(\mathbf{k}_{T}^{\prime}+k_{z}^{\prime}\hat{\mathbf{z}}\right)\times\left[\tilde{\mathbf{E}}^\text{i}\left(\mathbf{k}_{T}^{\prime},\omega\right)+\tilde{\mathbf{E}}^\text{t}\left(\mathbf{k}_{T}^{\prime},\omega\right)\right]+\left(\mathbf{k}_{T}^{\prime}-k_{z}^{\prime}\hat{\mathbf{z}}\right)\times\tilde{\mathbf{E}}^\text{r}\left(\mathbf{k}_{T}^{\prime},\omega\right)\right]d\mathbf{k}_{T}^{\prime},
\end{split} \label{eq:SD-IE-1}
\end{equation}
\end{figure*}
\begin{figure*}[!ht]
\begin{equation}
\begin{split}
&\mathbf{\hat{z}}\times\left(-\tilde{\mathbf{E}}^\text{i}\left(\mathbf{k}_{T},\omega\right)-\tilde{\mathbf{E}}^\text{r}\left(\mathbf{k}_{T},\omega\right)+\tilde{\mathbf{E}}^\text{t}\left(\mathbf{k}_{T},\omega\right)\right)= \\
&-\frac{1}{2}\mu_{0}i\omega\int\frac{1}{\omega\mu_0}\tilde{\bar{\bar{\boldsymbol{\chi}}}}_\text{mm}^T\left(\mathbf{k}_{T}-\mathbf{k}_{T}^{\prime},\omega\right)
\left[\left(\mathbf{k}_{T}^{\prime}+k_{z}^{\prime}\hat{\mathbf{z}}\right)\times\tilde{\mathbf{E}}^\text{i}\left(\mathbf{k}_{T}^{\prime},\omega\right)+\left(\mathbf{k}_{T}^{\prime}-k_{z}^{\prime}\hat{\mathbf{z}}\right)\times\tilde{\mathbf{E}}^\text{r}\left(\mathbf{k}_{T}^{\prime},\omega\right)+\left(\mathbf{k}_{T}^{\prime}+k_{z}^{\prime}\hat{\mathbf{z}}\right)\times\tilde{\mathbf{E}}^\text{t}\left(\mathbf{k}_{T}^{\prime},\omega\right)\right]d\mathbf{k}_{T}^{\prime} \\
&-\frac{1}{2\epsilon_{0}}\varepsilon_{0}\int\hat{\mathbf{z}}\times\left(\mathbf{k}_{T}^{\prime}-\mathbf{k}_{T}\right)\tilde{\bar{\bar{\boldsymbol{\chi}}}}_\text{ee}^{z}\left(\mathbf{k}_{T}^{\prime}-\mathbf{k}_{T},\omega\right)\left[\tilde{\mathbf{E}}^\text{i}\left(\mathbf{k}_{T}^{\prime},\omega\right)+\tilde{\mathbf{E}}^\text{r}\left(\mathbf{k}_{T}^{\prime},\omega\right)+\tilde{\mathbf{E}}^\text{t}\left(\mathbf{k}_{T}^{\prime},\omega\right)\right]d\mathbf{k}_{T}^{\prime}.
\end{split} \label{eq:SD-IE-2}
\end{equation}
\end{figure*}
\noindent
Equations~\eqref{eq:SD-IE-1} and \eqref{eq:SD-IE-2} represent four integral equations in six unknowns $\tilde{E}^\text{r}_x$, $\tilde{E}^\text{r}_y$, $\tilde{E}^\text{r}_z$, $\tilde{E}^\text{t}_x$, $\tilde{E}^\text{t}_y$, $\tilde{E}^\text{t}_z$. However, as $\mathbf{E}^\text{r}$ and $\mathbf{E}^\text{t}$ satisfy the free-space Maxwell equations, the normal components $\tilde{E}^\text{r}_z$ and $\tilde{E}^\text{t}_z$ are related to the transverse components through the divergence relations $\nabla\cdot\mathbf{E}^\text{r}(\mathbf{r}, t)=0$, $\nabla\cdot\mathbf{E}^\text{t}(\mathbf{r}, t)=0$ and are thus eliminated using
\begin{subequations}
\begin{align}
-j(\mathbf{k}_T+k_z\hat{\mathbf{z}})&\cdot\tilde{\mathbf{E}}^\text{t}=0 \nonumber \\
&\rightarrow \tilde{E}_z^\textbf{t} = \frac{-1}{k_z}(k_x\tilde{E}_x^\textbf{t}  +k_y\tilde{E}_y^\textbf{t}), \\
-j(\mathbf{k}_T-k_z\hat{\mathbf{z}})&\cdot\tilde{\mathbf{E}}^\text{r}=0 \nonumber \\
&\rightarrow \tilde{E}_z^\textbf{r} = \frac{1}{k_z}(k_x\tilde{E}_x^\textbf{r}  +k_y\tilde{E}_y^\textbf{r}).
\end{align} \label{eq:divE-kw-Ez}
\end{subequations}
Substituting~\eqref{eq:divE-kw-Ez} into~\eqref{eq:SD-IE-1} and~\eqref{eq:SD-IE-2} finally results in a system of four coupled integral equations in the four unknowns $\tilde{E}^\text{r}_x(\mathbf{k}_T, \omega)$, $\tilde{E}^\text{r}_y(\mathbf{k}_T, \omega)$, $\tilde{E}^\text{t}_x(\mathbf{k}_T, \omega)$, $\tilde{E}^\text{t}_y(\mathbf{k}_T, \omega)$, that may be solved by the method of moments.

\section{Integral-Equation System Resolution by the Method of Moments} \label{sec:MoM}

The SD integral-equation system~\eqref{eq:SD-IE-1} and \eqref{eq:SD-IE-2} with~\eqref{eq:divE-kw-Ez} are expressed in the temporal frequency ($\omega$) domain. However, if the system is linear, they naturally also apply to non time harmonic problems, since any (non time harmonic) field may be analyzed as the superposition of its time-harmonic Fourier components. In particular, a pulse incident field may be written as
\begin{equation}
\mathbf{E}^\text{i}\left(\boldsymbol{\rho},t\right)= \int_{-\infty}^{+\infty}\mathbf{E}^\text{i}\left(\boldsymbol{\rho},\omega\right)e^{j\omega t}d\omega.
\end{equation}
\noindent For this excitation, we first transform $\mathbf{E}^\text{i}\left(\boldsymbol{\rho},\omega\right)$, $\bar{\bar{\boldsymbol{\chi}}}_\text{ee}\left(\boldsymbol{\rho},\omega\right)$ and $\bar{\bar{\boldsymbol{\chi}}}_\text{mm}\left(\boldsymbol{\rho},\omega\right)$ into their spatial Fourier harmonics, namely $\tilde{\mathbf{E}}^\text{i}\left(\mathbf{k}_{T},\omega\right)$, $\tilde{\bar{\bar{\boldsymbol{\chi}}}}_\text{ee}\left(\mathbf{k}_{T},\omega\right)$ and $\tilde{\bar{\bar{\boldsymbol{\chi}}}}_\text{mm}\left(\mathbf{k}_{T},\omega\right)$, which are used as (known) inputs to the integral equation system. The system is then solved so as to determine the unknowns $\tilde{E}^\text{r}_x(\mathbf{k}_T, \omega)$, $\tilde{E}^\text{r}_y(\mathbf{k}_T, \omega)$, $\tilde{E}^\text{t}_x(\mathbf{k}_T, \omega)$ and $\tilde{E}^\text{t}_y(\mathbf{k}_T, \omega)$. The process is repeated for all the frequencies $\omega$ of the input wave\footnote{No new frequencies are created since the system is assumed to be both linear and time-invariant.} spectrum. Finally, the space-time dependent reflected and transmitted fields are found using \eqref{eq:E-kw}. In the case of a time-harmonic incident wave, the system is naturally solved only once, for the incident frequency.

For each frequency sample $\omega$, the SD integral-equation system~\eqref{eq:SD-IE-1} and~\eqref{eq:SD-IE-2} with~\eqref{eq:divE-kw-Ez} is solved using a standard method of moment (MoM) technique in the spectral $\mathbf{k}_T$ domain~\cite{harrington1996field, chamanara2013_aps_guidedIE,chamanara2013_iceaa_hybridguidedIE}. First, the unknowns are expanded over a set of basis functions, $B_n(\mathbf{k}_T)$, as
\begin{subequations}
\begin{align}
\tilde{E}^\text{r}_x(\mathbf{k}_T, \omega) = \sum a_n^{\text{r}x} B_n(\mathbf{k}_T) \\
\tilde{E}^\text{r}_y(\mathbf{k}_T, \omega) = \sum a_n^{\text{r}y} B_n(\mathbf{k}_T) \\
\tilde{E}^\text{t}_x(\mathbf{k}_T, \omega) = \sum a_n^{\text{t}x} B_n(\mathbf{k}_T) \\
\tilde{E}^\text{t}_y(\mathbf{k}_T, \omega) = \sum a_n^{\text{t}y} B_n(\mathbf{k}_T)
\end{align}
\end{subequations}
where $a_n^{\text{r}x}$, $a_n^{\text{r}y}$, $a_n^{\text{t}x}$, $a_n^{\text{t}y}$ represent unknown coefficients. Equations~\eqref{eq:SD-IE-1} and \eqref{eq:SD-IE-2} are then multiplied by a set of SD testing functions $W_n(\mathbf{k}_T)$ and integrated over the entire $\mathbf{k}_T$ domain. This process reduces \eqref{eq:SD-IE-1} and \eqref{eq:SD-IE-2} with~\eqref{eq:divE-kw-Ez} to a system of linear equations for the unknown coefficients $a_n^{\text{r}x}$, $a_n^{\text{r}y}$, $a_n^{\text{t}x}$, $a_n^{\text{t}y}$. In the forthcoming illustrative examples, we have used point matching, i.e. $W_n(\mathbf{k}_T)=\delta(\mathbf{k}_T - \mathbf{k}_{nT})$, and Galerkin testing, i.e. $B_n(\mathbf{k}_T)=W_n(\mathbf{k}_T)$.

\section{Illustrative Examples}   \label{sec:example}

This section presents two illustrative examples of application of the proposed SD integral-equation resolution technique. Section~\ref{sec:example-gen-refract} considers a monochromatic generalized-refractive metasurface, while Sec.~\ref{sec:example-poly-lens} considers a polychromatic focusing metasurface, with different focal points at different frequencies. In both cases, we first obtain the metasurface susceptibility functions for the specified transformation using the GSTC synthesis technique presented in~\cite{achouri2014general}, and next analyze the resulting metasurface using the SD integral-equation technique and verify that the synthesized operation is properly achieved.

\subsection{Monochromatic Generalized-Refractivie Metasurface} \label{sec:example-gen-refract}

The monochromatic generalized-refractive metasurface represented in Fig.~\ref{fig:pw-R-T}. It transforms a given incident plane wave with angle $\theta^\text{i}$ into the specified reflected and transmitted waves, with transmission and reflection coefficients $T$ and $R$ and angles $\theta^\text{r}$ and $\theta^\text{t}$, respectively. Assuming that the metasurface is monoisotropic and devoid of normal susceptibilities, its susceptibility tensors reduce to
\begin{subequations}
\begin{align}
\bar{\bar{\boldsymbol{\chi}}}_\text{ee}\left(\boldsymbol{\rho},\omega\right) &= \left[\begin{array}{ccc}
\chi_\text{ee}\left(\boldsymbol{\rho},\omega\right) & 0 & 0\\
0 & \chi_\text{ee}\left(\boldsymbol{\rho},\omega\right) & 0\\
0 & 0 & 0
\end{array}\right], \\
\bar{\bar{\boldsymbol{\chi}}}_\text{mm}\left(\boldsymbol{\rho},\omega\right) &= \left[\begin{array}{ccc}
\chi_\text{mm}\left(\boldsymbol{\rho},\omega\right) & 0 & 0\\
0 & \chi_\text{mm}\left(\boldsymbol{\rho},\omega\right) & 0\\
0 & 0 & 0
\end{array}\right],
\end{align}
\end{subequations}
corresponding to the tangential and normal polarization densities
\begin{subequations}
\begin{align}
\mathbf{P}_T\left(\boldsymbol{\rho},\omega\right) &= \chi_\text{ee}\left(\boldsymbol{\rho},\omega\right)\mathbf{E}_{\text{av},T}\left(\boldsymbol{\rho},\omega\right), \\
P_z\left(\boldsymbol{\rho},\omega\right) &= 0, \\
\mathbf{M}_T\left(\boldsymbol{\rho},\omega\right) &= \chi_\text{mm}\left(\boldsymbol{\rho},\omega\right)\mathbf{H}_{\text{av},T}\left(\boldsymbol{\rho},\omega\right), \\
M_z\left(\boldsymbol{\rho},\omega\right) &= 0.
\end{align} \label{eq:Xt-iso-Xz-0}
\end{subequations}

\begin{figure}[ht!]
\centering
\psfrag{a}[l][c][1.0]{$\mathbf{k}^\text{i}$}
\psfrag{b}[l][c][1.0]{$\mathbf{E}^\text{i}$}
\psfrag{c}[l][c][1.0]{$\mathbf{H}^\text{i}$}
\psfrag{d}[l][c][1.0]{$\mathbf{k}^\text{r}$}
\psfrag{e}[l][c][1.0]{$\mathbf{E}^\text{r}$}
\psfrag{f}[l][c][1.0]{$\mathbf{H}^\text{r}$}
\psfrag{g}[l][c][1.0]{$\mathbf{k}^\text{t}$}
\psfrag{h}[l][c][1.0]{$\mathbf{E}^\text{t}$}
\psfrag{i}[l][c][1.0]{$\mathbf{H}^\text{t}$}
\psfrag{j}[l][c][1.0]{$\theta^\text{i}$}
\psfrag{k}[l][c][1.0]{$\theta^\text{r}$}
\psfrag{l}[l][c][1.0]{$\theta^\text{t}$}
\psfrag{y}[l][c][0.9]{$y$}
\psfrag{z}[l][c][0.9]{$z$}
\psfrag{m}[l][c][0.9]{metasurface}
\includegraphics[page=2, width=0.9\columnwidth]{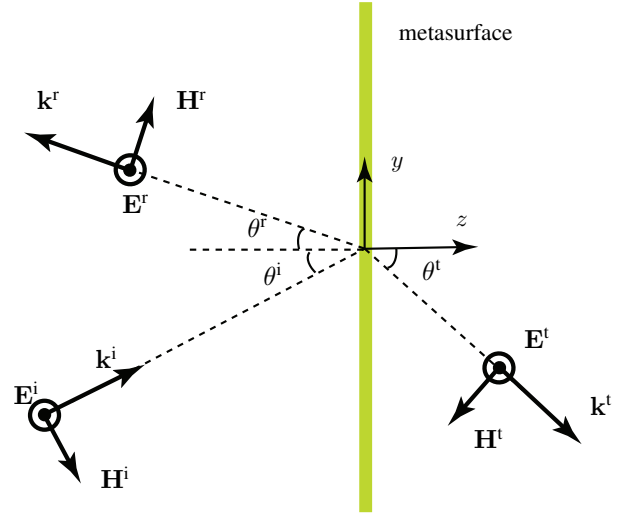}
\caption{Monochromatic generalized-refractive metasurface. The incident plane wave is reflected and transmitted with specified reflection and transmission coefficients and angles.}
\label{fig:pw-R-T}
\end{figure}

\begin{figure}[ht!]
\subfigure[~~~~~~~~~~~~~~~~~~~~~~~~~~~~~~~~~~~~~~~~~~~~~~~~~~~~~~~~~~~~~~~~~~~~~~~~~~~~~~~~~~~~~~~~~~~]{\label{fig:pw-Xee}
\includegraphics[page=3, width=0.85\columnwidth]{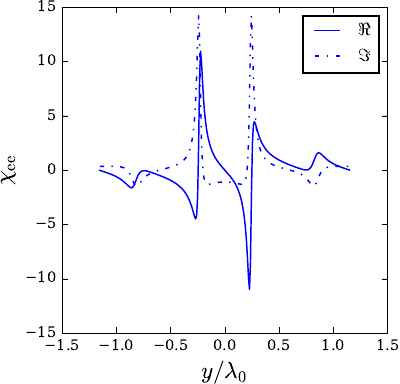}}
\subfigure[~~~~~~~~~~~~~~~~~~~~~~~~~~~~~~~~~~~~~~~~~~~~~~~~~~~~~~~~~~~~~~~~~~~~~~~~~~~~~~~~~~~~~~~~~~~]{\label{fig:pw-Xmm}
\includegraphics[page=4, width=0.85\columnwidth]{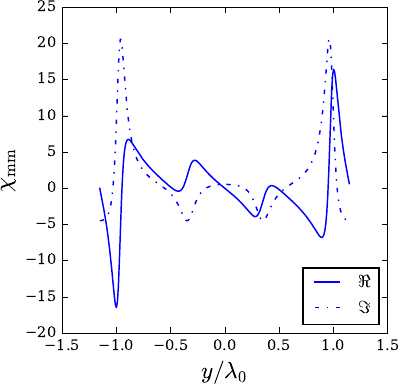}}
\caption{Synthesized electric and magnetic susceptibilities for the generalized refraction problem in Fig.~\ref{fig:pw-R-T} with $\theta^\text{i}=60^\circ$, $\theta^\text{r}=25^\circ$, $\theta^\text{t}=60^\circ$, $T=0.8$ and $R=-0.6$.}
\label{fig:pw-X}
\end{figure}

We require that the incident wave impinge on the metasurface with the angle $\theta^i=60^\circ$, negatively refract with the angle $\theta^t=60^\circ$ and transmission coefficient $T=0.8$, and reflect with the angle $\theta^r=25\deg$ and reflection coefficient $R=-0.6$. The metasurface satisfying to these specifications is readily synthesized from the GSTCs relations

\begin{subequations}
\begin{align}
\chi_\text{ee}\left(y,\omega\right) &= \Delta H_y\left(y,\omega\right)/(j\omega\epsilon_0 E_{\text{av},x}\left(y,\omega\right)), \\
\chi_\text{mm}\left(y,\omega\right) &= \Delta E_x\left(y,\omega\right)/(j\omega\mu_0 H_{\text{av},y}\left(y,\omega\right)),
\end{align}
\end{subequations}

\noindent where the difference and average fields are composed of the specified fields according to~\eqref{eq:diff_fields} and~\eqref{eq:av_fields}. Since the specified fields have no $x$ dependence, the susceptibilities are only functions of $y$. They are plotted in Fig.~\ref{fig:pw-X}. Note that as this transformation conserves the total power $|T|^2+|R|^2=1$, losses in the susceptibilities should be compensated by gains. Therefore in some region the imaginary part of susceptibilities become positive, corresponding to gain.

From this point, we shall carry out the analysis and see if we thereby retrieve the synthesized operation. Given the incident field and synthesized metasurface susceptibilities, we compute the reflected and transmitted fields using the SD integral-equation and MoM technique. The results are plotted in Fig.~\ref{fig:pw-R-T}. They exactly correspond to the specified fields, which validates the analysis technique.

\begin{figure}[ht!]
\includegraphics[page=5, width=1.0\columnwidth]{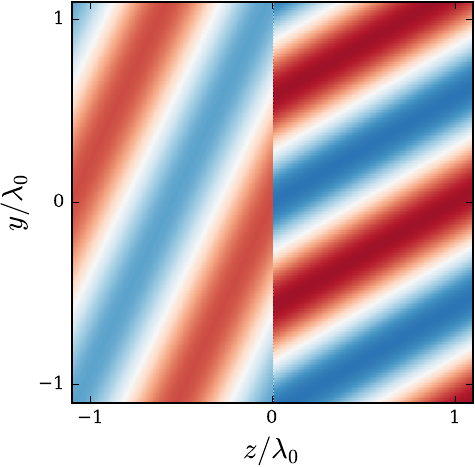}
\caption{Reflected and transmitted fields computed by the SD integral-equation and MoM technique for the metasurface susceptibilities in Fig.~\ref{fig:pw-X}.}
\label{fig:pw-R-T}
\end{figure}

\subsection{Polychromatic Focusing Metasurface}  \label{sec:example-poly-lens}

\begin{figure}[ht!]
\centering
\psfrag{a}[l][c][1.0]{$\omega_1$}
\psfrag{b}[l][c][1.0]{$\omega_2$}
\psfrag{c}[l][c][1.0]{$\omega_3$}
\psfrag{d}[l][c][1.0]{$\omega_4$}
\psfrag{e}[l][c][1.0]{$\omega_1$}
\psfrag{f}[l][c][1.0]{$\omega_2$}
\psfrag{g}[l][c][1.0]{$\omega_3$}
\psfrag{h}[l][c][1.0]{$\omega_4$}
\psfrag{y}[l][c][0.9]{$y$}
\psfrag{z}[l][c][0.9]{$z$}
\includegraphics[page=6, width=1.0\columnwidth]{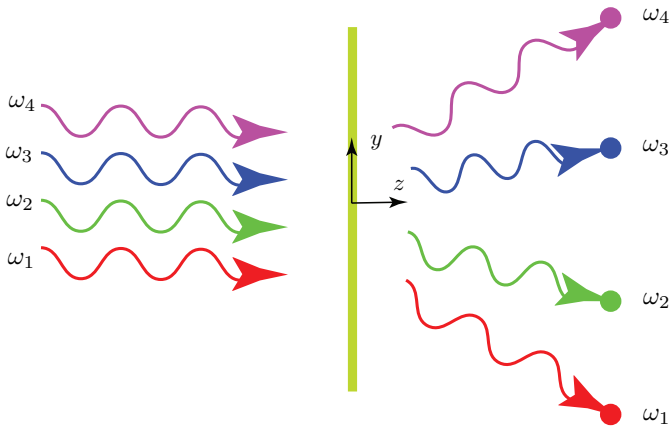}
\caption{Polychromatic focusing metasurface operating as a real-time spectrum analyzer: the different frequency contents of an input plane wave pulse are focused at different focal points in real time.}
\label{fig:bb-ms-rtsa}
\end{figure}

The polychromatic focusing metasurface is designed so as to exhibit different focal points at different frequencies, as shown in~Fig.~\ref{fig:bb-ms-rtsa}. A plane wave multi-wavelength pulse normally impinges on the metasurface from the left, and the metasurfaces focuses the different wavelengths or frequencies onto corresponding different points on the right. Such a metasurface essentially operates as a real-time spectrum analyzer as it spatially decomposes the frequency contents of the input pulse in real time. To synthesize the metasurface, we make separate designs for each of the frequencies $\omega_i$ and calculate the corresponding susceptibilities $\chi_\text{ee}\left(y,\omega_i\right)$, $\chi_\text{mm}\left(y,\omega_i\right)$ from the specified electromagnetic fields at the left and right sides of the metasurface.

Consider the specified field on the left side of the metasurface to be a $y$-polarized normally incident plane wave with frequency $\omega_i$ designed to create a $y$-polarized focus at the corresponding specified focal point. The required field on the right side of the metasurface can be calculated by placing a $y$-polarized dipole at the frequency $\omega_i$ on the focal point, calculating the field radiated by this dipole on the metasurface at $z=0_+$ and applying phase conjugation to this field to achieve reverse propagation. The susceptibilities $\chi_\text{ee}\left(y,\omega_i\right)$, $\chi_\text{mm}\left(y,\omega_i\right)$ are then computed as in Sec.~\ref{sec:example-gen-refract}. Again, as in \eqref{eq:Xt-iso-Xz-0}, we assume purely transverse and monoisotropic susceptibilities.

\begin{figure}[ht!]
\subfigure[~~~~~~~~~~~~~~~~~~~~~~~~~~~~~~~~~~~~~~~~~~~~~~~~~~~~~~~~~~~~~~~~~~~~~~~~~~~~~~~~~~~~~~~~~~~]{\label{fig:bb-Xee}
\includegraphics[page=7, width=0.85\columnwidth]{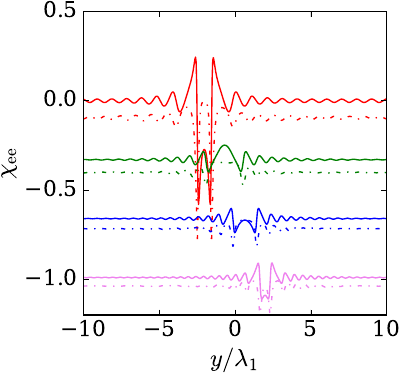}}
\subfigure[~~~~~~~~~~~~~~~~~~~~~~~~~~~~~~~~~~~~~~~~~~~~~~~~~~~~~~~~~~~~~~~~~~~~~~~~~~~~~~~~~~~~~~~~~~~]{\label{fig:bb-Xmm}
\includegraphics[page=8, width=0.85\columnwidth]{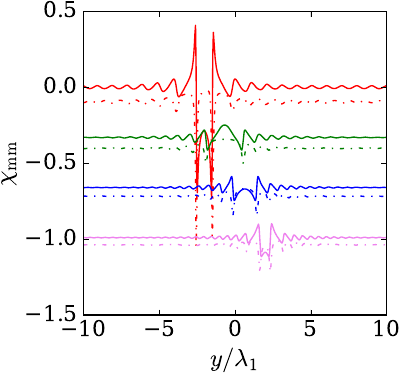}}
\caption{Electric and magnetic susceptibilities corresponding to the polychromatic focusing metasurface in Fig.~\ref{fig:bb-ms-rtsa}. The solid and dashed curves corresponding to real and imaginary parts, respectively. For clarity the green, blue and violet curves are vertically offset by $-0.33$, $-0.66$, $-1.0$ units, respectively.}
\label{fig:bb-X}
\end{figure}

The so obtained susceptibilities are plotted in Fig.~\ref{fig:bb-X} for a four-wavelength metasurface with equidistant frequencies across the range $\omega_1\cdots2\omega_1$ and corresponding four focal points. The horizontal axis is normalized to the wavelength at frequency $\omega_1$. The focal points are separated by $1.33\lambda_1$ from $-2\lambda_1$ to $2\lambda_1$. Note that the susceptibilities exhibit maxima at the points corresponding to their respective focal height. The electric and magnetic susceptibilities are almost identical, corresponding to a matched, and hence reflection-less, metasurface~\cite{gupta2016perfect}.

\begin{figure}[ht!]
\subfigure[~~~~~~~~~~~~~~~~~~~~~~~~~~~~~~~~~~~~~~~~~~~~~~~~~~~~~~~~~~~~~~~~~~~~~~~~~~~~~~~~~~~~~~~~~~~]{\label{fig:bb-Ey-mag}
\includegraphics[page=9, width=1.0\columnwidth]{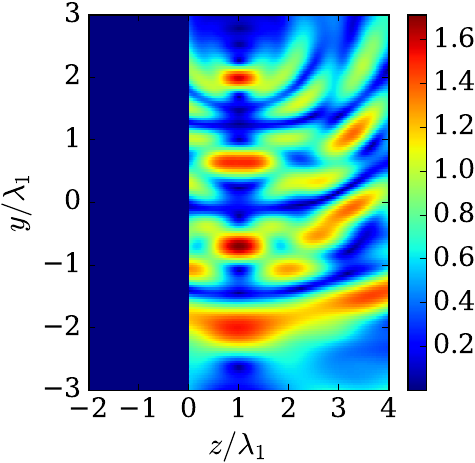}}
\subfigure[~~~~~~~~~~~~~~~~~~~~~~~~~~~~~~~~~~~~~~~~~~~~~~~~~~~~~~~~~~~~~~~~~~~~~~~~~~~~~~~~~~~~~~~~~~~]{\label{fig:bb-Ey_phase}
\includegraphics[page=10, width=1.0\columnwidth]{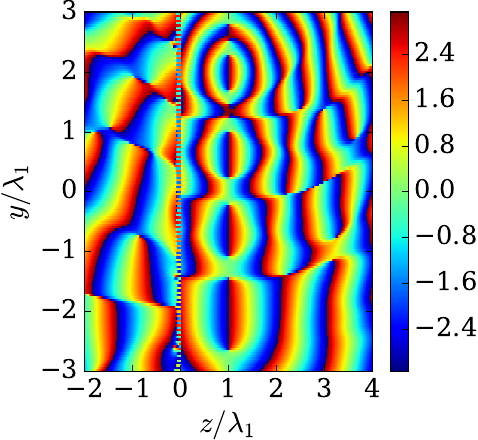}}
\caption{Sum of reflected and transmitted fields $E_y$ at $\omega_1$ to $\omega_4$ computed using the SD integral-equation and MoM technique for the metasurface susceptibilities in Fig.~\ref{fig:bb-X}. (a)~Magnitude. (b)~Phase.}
\label{fig:bb-Ey-RT}
\end{figure}

\begin{figure}[ht!]
\subfigure[~~~~~~~~~~~~~~~~~~~~~~~~~~~~~~~~~~~~~~~~~~~~]{\label{fig:bb-Ey-mag-1}
\includegraphics[page=11, width=0.45\columnwidth]{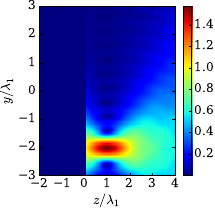}}
\hfill
\subfigure[~~~~~~~~~~~~~~~~~~~~~~~~~~~~~~~~~~~~~~~~~~~~]{\label{fig:bb-Ey-mag-2}
\includegraphics[page=12, width=0.45\columnwidth]{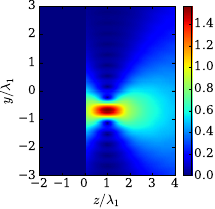}}
\subfigure[~~~~~~~~~~~~~~~~~~~~~~~~~~~~~~~~~~~~~~~~~~~~]{\label{fig:bb-Ey-mag-3}
\includegraphics[page=13, width=0.45\columnwidth]{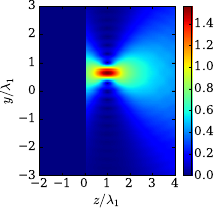}}
\hfill
\subfigure[~~~~~~~~~~~~~~~~~~~~~~~~~~~~~~~~~~~~~~~~~~~~]{\label{fig:bb-Ey-mag-4}
\includegraphics[page=14, width=0.45\columnwidth]{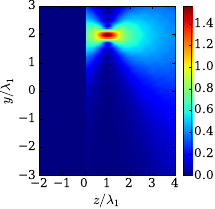}}
\caption{Separate magnitudes of the reflected and transmitted electric fields in Fig.~\ref{fig:bb-Ey-RT} (a)~at $\omega_1$, (b)~at $\omega_2$, (c)~at $\omega_3$, and (d)~at $\omega_4$.}
\label{fig:bb-Ey-RT-14}
\end{figure}

Next we perform the SD integral-equation and MoM analysis of the metasurface just synthesized. The sum of the resulting reflected and transmitted fields at frequencies $\omega_1$ to $\omega_4$ are plotted in Fig.~\ref{fig:bb-Ey-RT}. The metasurface is well matched (no reflection) and generates focal spots at the four specified points. Moreover, these spots correspond to the specified frequencies as may deduced from the corresponding wavelengths. The focal spots do not seem very accurate in Fig.~\ref{fig:bb-Ey-RT}, but this is due to the superposition of the four waves of frequencies $\omega_1$ to $\omega_4$ and this does not represent a practical issue since the different frequencies are orthogonal and hence independent from each other. Figure~\ref{fig:bb-Ey-RT-14}, that shows the response of the metasurface separately excited by the four waves, confirms this fact.

\section{Conclusion}	\label{sec:conclusion}
We have presented a spectral domain surface integral equation technique for the analysis of metasurfaces. A system of four surface integral equations have been derived for the reflected and transmitted electric fields, and solved using the method of moments. Compared to the finite difference and finite element techniques that require volumetric meshes, the proposed technique reduces the problem to the surface of the metasurface, eliminating one dimension and therefore providing benefits in terms of memory and speed. A monochromatic generalized refraction metasurface and a polychromatic focusing metasurface have been presented as illustrative examples.


%

%
%

\ifCLASSOPTIONcaptionsoff
  \newpage
\fi



%
%
%

\bibliographystyle{IEEEtran}
\bibliography{ReferenceList2_abbr}

%

%
%
%




\end{document}